# Developing triangular-shaped tiny particles and mono-layer shapes at the air-solution interface, and binding of mono-layers into a particle

**Mubarak Ali**

Islamabad, Park Road, Pakistan, mubarak3974@gmail.com, mubarak74@mail.com, mubarak60@hotmail.com, https://www.researchgate.net/profile/Mubarak-Ali-7, http://orcid.org/0000-0003-1612-6014

**Abstract:** The new insights into the atomic structure empower one to develop materials by a bottom-up approach. The study of colloids is profitable in many ways. Considering the research in this field is beneficial. Developing tiny metallic particles in a specific shape is a need. It is a need for many cutting-edge applications. The different steps involved in the development of triangular-shaped tiny particles, mono-layer shapes, nanoparticles, and particles are not yet clear. In processing solutions by different means, atoms should first dissociate from the precursor. The pulses of nano energy bind the atoms of the monolayered assembly. There can be other sources of a nanoenergy packet binding atoms. Triangular-shaped tiny particles act as the building blocks of later-developed mono-layer shapes, nanoparticles and particles. Triangular-shaped tiny particles leave the electronically flat solution surface to enter the electronically decreasing solution surface. Arrays of tiny particles convert into structures of smooth elements. The structures of smooth elements develop a mono-layer shape. It is at the center of the concave meniscus. At the air-solution interface, traveling photons further flatten the structures of the smooth elements, developing a mono-layer shape.

The gravitational force at the electron level in the upper mono-layer shape is greater than the levitational force. The opposite is for a lower mono-layer shape. The time to adhere two mono-layer shapes is only a few microseconds. This study discusses the developments of triangular-shaped tiny particles and mono-layer shapes at the air-solution interface. It further explores the binding of two mono-layer shapes. A geometric nanoparticle or particle develops. The current study has great worth in nanoscience and nanotechnology. It also boosts other fields of science, engineering, and technology.

## 1.0. Introduction

Synthesizing materials for some planned applications has always been crucial. Controlling the size and shape of materials is at the frontier of material science, physics, chemistry, and nanoscience. Due to the decrease in the scale of materials, new perspectives are also emerging. These kinds of opportunities can open new avenues for thinking and gaining knowledge.

The development of morphological structure in various materials is a promising field. Geometric and anisotropic objects can lead to many unique applications. Some applications can also have a social impact. Others have an impact on the medical fields. Particles of different shapes have their usual applications in the many fields of science and engineering. It is possible to observe the structure of materials at the nanoscale and atomic scale.

A microscopic study of the material at high magnification is feasible for a long time. The roles of dynamics remained in illusion. Many high-resolution microscopy images in the scientific literature can reveal the dynamics of tiny particles and their atoms. Nanoparticles and particles amalgamated in solution under different driving forces [1]. In the development of geometric particles, the role of the atomic dynamics is crucial, as discussed elsewhere [2-5]. A greater field emission was due to the graphitic phase of the carbon film, as discussed elsewhere [6].

The switching dynamics of carbon atoms deposited the carbon films with different morphologies and structures [7]. Photons of suitable lengths and numbers shape the lattice of an atomic element, as discussed elsewhere [8]. Atoms dissociate from precursors when heat is supplied [9]. Rather than revealing negative or positive charge, atoms adopt other options as discussed elsewhere [10]. The electron dynamics of the silicon atom convert heat energy into photon energy. A study elsewhere [11] discusses this. Suitable element atoms execute confined interstate electron dynamics to generate binding energy [12]. Tiny clusters of gold behave like simple chemical compounds [13]. Nanocrystals can develop higher-order materials due to their unique features [14].

Upon trapping energetic electrons, tiny particles oscillate collectively [15]. To fabricate small electronic devices, nanoparticle-based technology has the potential [16]. The assembly of tiny particles shape a nanoparticle or particle [17]. A study elsewhere [18] focuses on assembling nanoparticles in a higher-order functional material. A study elsewhere [19] demonstrates the computer simulation of a unique self-assembled structure. To achieve a successful assembly, nanoparticles and colloidal building blocks act as the atoms and molecules of future materials [20]. Understanding dynamics is essential for developing nanoparticles into ordered arrays [21]. Precise control of the surface properties of nanoparticles helps in the design of assemblies with higher-order structures [22].

Tiny particles are molecular-like structures defining hexagonal close-packed (hcp) structures [23]. In addition to the structure, the dynamics also discuss entropy and geometry [24]. To characterize the order of packing, some metrics are also capable, as discussed elsewhere [25]. Many studies on plasma solution technologies have discussed the development of nanoparticles and particles. The synthesis of nanoparticles revealed different discharge characteristics [26]. At a constant input current, the electron flux remains the same on the solution surface [27].

To lower the solution pH, plasma electrons convert H radicals into molecular hydrogen [28]. Brownian motion and surface charge both explain the stability of nanoparticles, as discussed elsewhere [29]. Several physical and chemical aspects contribute to the development of nanomaterials at interfaces [30]. The concentration of electrons is controllable under the plasma in contact with a solution [31]. Numerous studies discussed the developmental mechanisms of tiny particles.

This study discusses the fundamental process of developing triangular-shaped tiny particles and mono-layer shapes. A mono-layer shape can also refer to a single atomic layer, a thin particle layer, or a geometrically constrained structure. The gold atoms explain the models. Triangular-shaped tiny particles develop a mono-layer shape. Two mono-layer shapes bind laterally to develop a geometric particle. The binding mechanisms discussed here are not available in the published literature.

The study discusses controlling the size and shape of the particles in processing solutions. It is relevant to a new atomic structure. This study is novel in this sense. In processing different colloidal solutions, the input power to the system is a photonic current. It is not an electric current. (The study in reference 10 provides further details about this.) The study is original in terms of the actual atomic structure. The study establishes a new ground for colloidal science and physical chemistry.

## 2.0. Experimental details/ Theoretical framework

Different models in this study support the experimental results, discussing the development of triangular-shaped tiny particles, mono-layer shapes, nanoparticles, and particles. Several methods for the processing of colloidal solutions are available in the literature. An electrochemical or a photochemical process is the most widely used method. Different-sized particles can also develop from a pulse-based electron-photon/solution interface process. For the development of different-sized particles, a solution-based process involving so-called ‘plasma’ is also suitable.

Several studies published elsewhere [1-5, 32, and 33] discussed the experimental details. The studies also discussed various-sized particles. The colloidal solutions of those developed different nano- and micro-sized particles processed by a pulse-based electron-photon/solution interface process. A pulse-based process is a new method. It is a robust method. The mono-layer shapes and the particles do not have adverse effects. Moreover, it is scalable and sustainable.

The published studies show many microscopic images of nanoparticles and particles. The reference studies 2 to 5 and 32 show the shapes of those nanoparticles and particles. Thus, the question is not about the experimental setup or method. The question is about the structure of atoms. Atoms develop a nanoparticle or particle. A study elsewhere [10] discusses the input power. A solution processing is due to the power. How atoms dissociate from their precursors is a key question.

To process different colloidal solutions, how does the reduction rate of the precursor contribute? How did triangular-shaped tiny particles develop from a monolayered assembly? How does a tiny particle develop a mono-layer shape of different geometrical particles? How does a mono-layer shape develop in particles of different shapes? These are a few questions. There are many more questions.

These questions are fundamental to physical and chemical sciences. These questions govern the science behind the development of nanoparticles and particles having different shapes. These questions can facilitate improving the design of a method for processing colloidal solutions. These questions also lead to the development of a specific shape of a nanoparticle or particle. There is also a need to address many remaining questions.

A microscope can show different shapes of nanoparticles and particles in a processed solution of a suitable molar concentration. In addition to experimental validation, theoretical, structural, and contextual engagements with the existing literature, the developed shapes would require a discussion of their underlying science. The existing study discusses the fundamental science and engineering behind different nanoparticles and particles. To show microscopic images of different nanoparticles and particles is not the subject of this study. The study by any solution-based process can benefit from the scientific and engineering details of this study.

The existing study can provide a foundation for future research. The published investigations do not address many underlying phenomena. If some studies discussed the underlying phenomena, there is speculation or confusing discussion. At many points, the earlier investigations are not sufficient. Thus, the target of this study is not to show the experimental setup or microscopic images.

The existing study builds a basic and applied concept underlying the development of various nanoparticles and particles. The reduction rate of a precursor largely varies the structural formation. Many experimental studies are needed to determine it. The force exertion is becoming increasingly non-uniform. The porosity also affects the structure. The setup parameters can also influence the dynamics of the solution processing. There is also a need to define several basic terms.

There is a mono-layer term when a tiny particle, nanoparticle, or particle develops with a single layer. The dynamics relate to the changing aspects of a matter at any scale. In the processing of colloids, the dynamics of tiny particles, mono-layer shapes, nanoparticles, and particles also influence the process. Only a few atoms bind to develop a tiny particle (or a tiny cluster). Binding mono-layer shapes to develop a nanoparticle or particle is also a part of the study.

The interaction of the light, photon, or glow, so-called plasma, with the solution also becomes the cause of dynamics. That interaction is also a part of the study of dynamics. Dynamics deal with the overall synergy of a processing solution. The interface between the air and the beaker solution is related to the air-solution interface. A surface force relates to the forces of the east and west poles.

A nanoenergy packet is a pulse. A pulse energy binds atoms of a monolayered assembly. Local gravity relates to the gravity that appears at the local level. Local gravity or local levity studies the local environment of the experimental setup. In some cases, local gravity (or levity) is a fundamental or universal gravity (or levity). A force of the east and west poles is a collective outcome of east-west forces.

## 3.0. Models and discussion

The hexagonal close-packed (hcp) structure should relate to a multidimensional shape. However, it refers to a two-dimensional structure in the earlier studies. The key point is to discuss the development of a triangular-shaped tiny particle. The development of tiny particles should be either in gold atoms or in atoms of other suitable elements. The present study addresses this issue. In the natural way of binding, atoms of suitable elements should evolve structures. In this case, the exertion of forces should remain in the relevant format, as discussed elsewhere [12].

The solid element atoms occupy the ground points below the ground surface. A grounded format exists below the ground surface. In this case, the force of the south pole remains for the solid atom electrons. Therefore, in the natural protocol or approach, atoms of suitable elements should evolve their structures. If the exertion of force is in the surface format, solid atoms can also develop their structures. It is due to the artificial protocol. In this case, solid atoms undergo their transition states. In the development of triangular-shaped tiny particles, the supplied controlled pulses bind the atoms. The studies elsewhere [2-5, 9] provide some initial information.

The binding between atoms consists of three types of forces. The Keesom force is the force between permanent dipoles. The Debye force is the force between a permanent dipole and an induced dipole. The London dispersion force is the force between induced dipoles. The binding between the atoms also does not exist due to the positive and negative charges. Moreover, atoms do not bind by electrostatic interactions.

A study elsewhere [10] reported that atoms do not form positive or negative charges. Furthermore, the binding of atoms also does not comply with the concepts of Bravais lattices. A study elsewhere [12] further discusses the non-validation of the concepts of Bravais lattices. Such conventional studies stop thinking forward. The binding of atoms should partially follow the concept of van der Waals forces. The energy should also contribute. By involving or engaging, the force should define its nature. In the development of distorted tiny particles, and depending on the energy-force details, dynamics contribute to binding atoms [33].

### *3.1. Development of triangular-shaped tiny particles*

Atoms dissociate from their precursors before levitating to the solution surface. On reaching the solution surface, atoms attain a required transition state. It is before the binding of the atoms into a monolayered assembly. A monolayered assembly of atoms is a single-layer assembly. Supplied nanoenergy packets bind atoms of the monolayered assembly. The tiny particles develop in a triangular shape.

At an electronically flat solution surface, the connected triangular-shaped tiny particles in blocks would develop. It is the case of a bipolar pulse. Before binding to a triangular-shaped tiny particle, atoms undergo a transition state at the solution surface. The transition state of the atoms can be related to a re-crystallization state. Most likely, the gold atoms reach a re-crystallized state during dissociation and levitation. However, additional research is needed to validate these findings.

A supplied nanoenergy packet develops a block of connected tiny particles. Each tiny particle developed in a triangular shape. It is due to the shape of the bipolar pulse. Figure 1 (a) shows this. Two triangular-shaped tiny particles join in each block. The size of a triangular-shaped tiny particle depends on the precursor concentration. With the unipolar mode of the pulse, a triangular-shaped tiny particle can develop directly. Figure 1 (b) shows this.

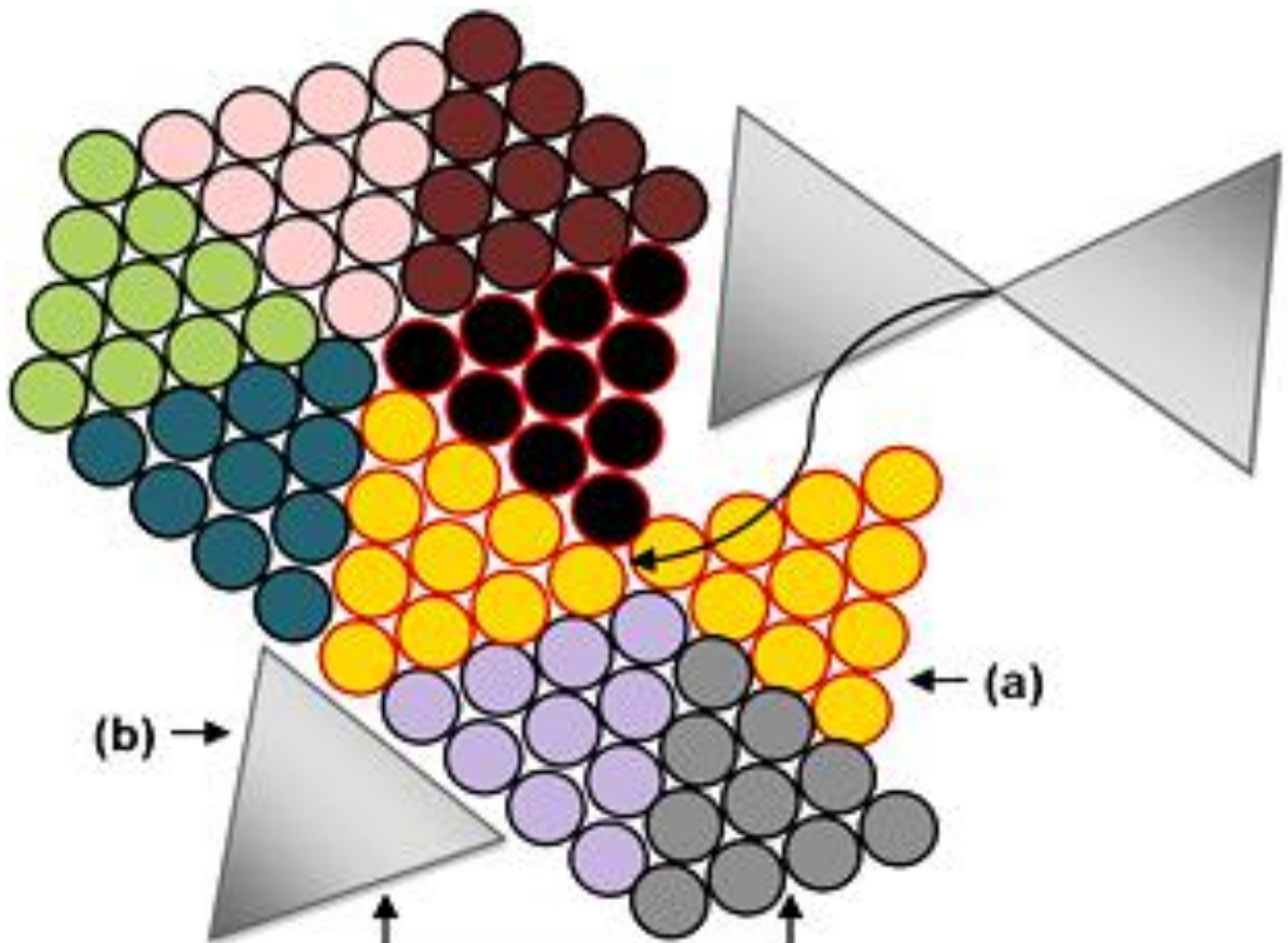

**Figure 1:** Tiny particles developing in the shape of nanoenergy packets from a monolayered assembly of atoms; (a) connected triangular-shaped tiny particles developing under the bipolar mode of pulse, and (b) a triangular-shaped tiny particle developing under the unipolar mode of pulse

A monolayered assembly of gold atoms develop at the solution surface. Figure 2 ($a_1$) shows this. Triangular-shaped tiny particles develop in a block. There is a bipolar pulse.

Two triangular-shaped tiny particles join in this case. Figure 2 ($a_2$) shows this. There is also a directly developed triangular-shaped tiny particle. Figure 2 ($a_3$) shows this. It is larger than the one shown in Figure 1 (b). Typically, tiny particles develop in the shape and size of the supplied nanoenergy packets.

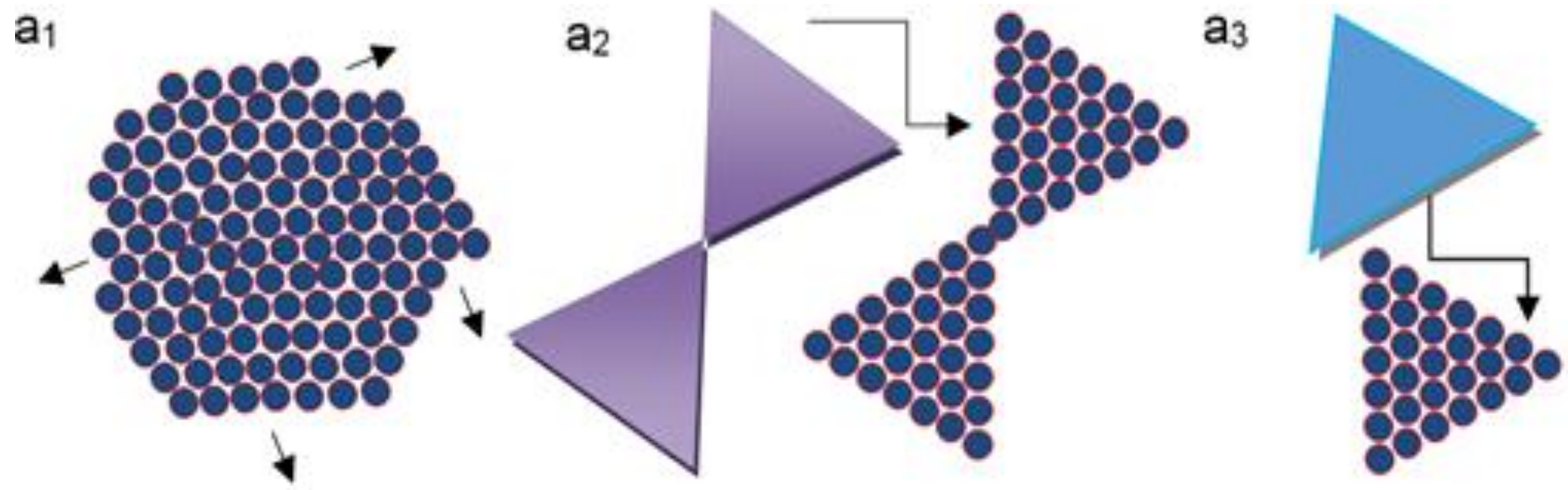

**Figure 2:** ($a_1$) Assembly of gold atoms at an electronically flat solution surface, ($a_2$) two connected triangular-shaped tiny particles, and ($a_3$) a single triangular-shaped tiny particle

It is also possible to achieve (or generate) packets of nanoenergy from other suitable sources. When packets of nanoenergy bind atoms, transitional atoms are ideally in their re-crystallized state. It is also feasible to develop building blocks in other shapes. The triangular-shaped tiny particles act as the building blocks in the development of a geometric nanoparticle (or particle). In earlier studies, they are the hcp structures.

### *3.2. Disconnecting connected triangular-shaped tiny particles*

The connected triangular-shaped tiny particles in Figure 3 (a) show the black dot. The black dot in Figure 3 (a) denotes the connecting point. The dot is between the horizontally bound atoms. Labels (1) and (2) in Figure 3 (a) denote the center of each horizontally positioned atom and the binding point of two tiny particles, respectively. The atoms positioned in the arrays of each triangular-shaped tiny particle behave differently in terms of force. The horizontally bound atoms connecting triangular-shaped tiny particles have a different force at the electron level.

The connected triangular-shaped tiny particles leave the electronically flat solution surface to enter the electronically decreasing solution surface. On entering an electronically decreasing solution surface, connected triangular-shaped tiny particles disconnected instantaneously. Connected tiny particles divide into two triangular-shaped tiny particles. Figure 3 (b) shows this. On entering the electronically decreasing solution surface, a force difference develops between the horizontally bound atoms of two connected triangular-shaped particles. From the point of connection of their horizontally bound atoms, two connected triangular-shaped tiny particles will disconnect.

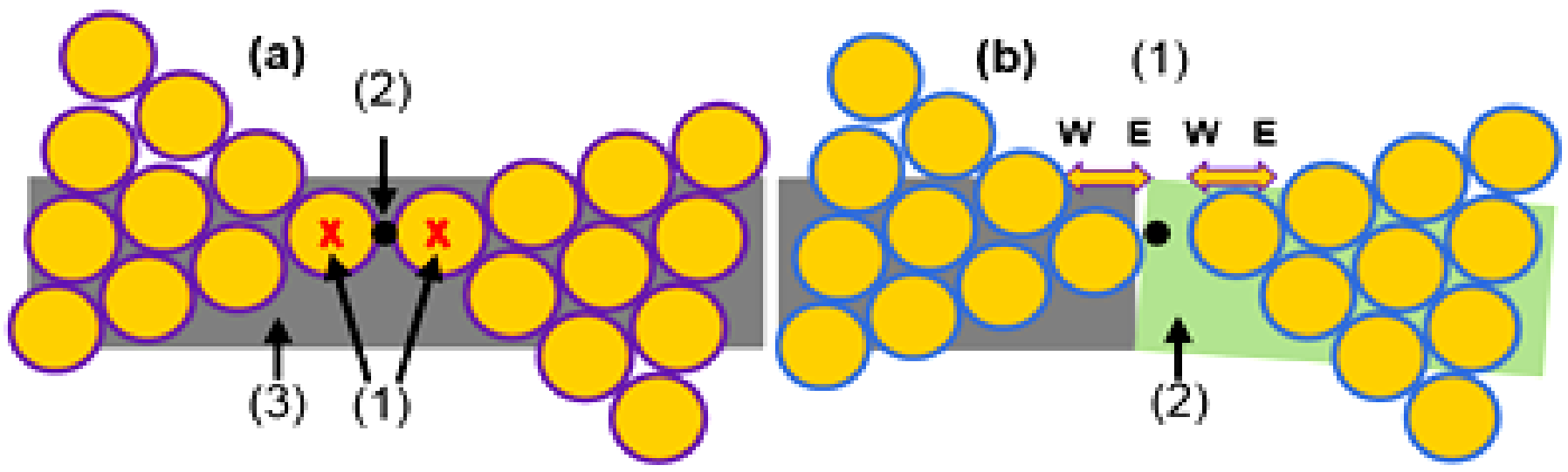

**Figure 3:** (a) Connected two triangular-shaped tiny particles in a block; (1) center of each horizontally bound atom, (2) connecting point of two horizontally bound atoms, and (3) electronically flat solution surface. (b) Disconnecting connected triangular-shaped tiny particles into two triangular-shaped tiny particles; (1) east (E) and west (W) poles of horizontally bound atoms, and (2) an electronically decreasing solution surface

The east-west poles of the horizontally bound atom are different. Label (1) in Figure 3 (b) denotes this. Label (2) in Figure 3 (b) denotes the disconnecting point between tiny particles, which is the boundary between the electronically flat and electronically decreasing solution surfaces. Different background colors indicate different solution surfaces. Figure 3 (b) also shows this. Two triangular-shaped tiny particles will result.

In the electronically decreasing solution surface, the two horizontally bound atoms (of connected triangular-shaped tiny particles) are no longer at the electronically flat solution surface. In the electronically decreasing solution surface, electrons of horizontally bound atoms do not experience a constant force from all four poles. The binding in the horizontally bound atoms is not like the binding in an array of atoms. To enter the electronically decreasing solution surface, the horizontally bound atoms at the connecting point of the triangular-shaped tiny particles disregard the adhesiveness.

At the connecting point of the triangular-shaped tiny particles, there is a disconnect in two horizontally bound atoms of the triangular-shaped tiny particles. The separated tiny particles deal with their dynamics separately. A slight difference between east and west forces (or north and south forces) develops a force difference at the electron level. The force difference at the electron level in 'horizontally bound atoms' causes the separation of two triangular-shaped tiny particles.

### *3.3. Elongation of atoms in arrays of a triangular-shaped tiny particle*

The elongation of a gold atom under the influence of even and uneven forces has been discussed elsewhere [10]. Atoms of transitional behavior deal with the immersing force at an electronically decreasing-level solution surface, as discussed elsewhere [32]. Thus, upon leaving the electronically flat solution surface and entering the electronically decreasing solution surface, the atoms in each array of a triangular-shaped tiny particle elongate. Depending on the conditions of a process, the elongation rate of the atoms in the triangular-shaped tiny particles varies.

The atoms in the arrays (of a triangular-shaped tiny particle) elongate at a lower rate. There is a decreased stretching rate of the atomic lattices. Figure 4 (a) shows this. In different arrays of a triangular-shaped tiny particle, the elongation of atoms is at a higher rate. Figure 4 (b) shows this. Atoms undergo transitional behavior at the same rate. Figures 4 (a) and (b) show this separately. In an array, the atoms adhere to each other during elongation. The decreased atomic elongation occurs in the arrays. Figure 4 (a) shows this. The increased atomic elongation occurs in the arrays. Figure 4 (b) shows this. In the elongation process, an atom binds with the lower-positioned atom.

Figures 4 (a) and (b) also show the single-atom elongation under the competing action of forces at opposite ends of their electrons. A study elsewhere [10] discussed the elongation of the single atom. However, the atomic elongation was on the flat surface. In each array of atoms, the supplied energy keeps their positions intact. At an electronically decreasing solution surface, though the atoms of each array compete with the forces of an immersing medium. The electrons tilt on both sides of the centre of an atom. The electron tilting safeguards the position of its atom in the array. In a triangular-shaped tiny particle, each atom of the array secures its position similarly.

In Figure 4 (c), label (1) denotes the electronically flat solution surface. In an electronically decreasing solution surface, the solution level decreased at the electron level. Label (2) in Figure 4 (c) shows this. Upon arrival at the concave meniscus, atomic arrays convert into the structures of the smooth elements. At a concave meniscus (developing at the solution surface) in a pulse-based system, each array of atoms has a shape like a smooth element. Label (3) in Figure 4 (c) shows the concave meniscus. Labels (4) and (5) in Figure 4 (c) indicate the tube carrying the glow sources and the glow spot, respectively.

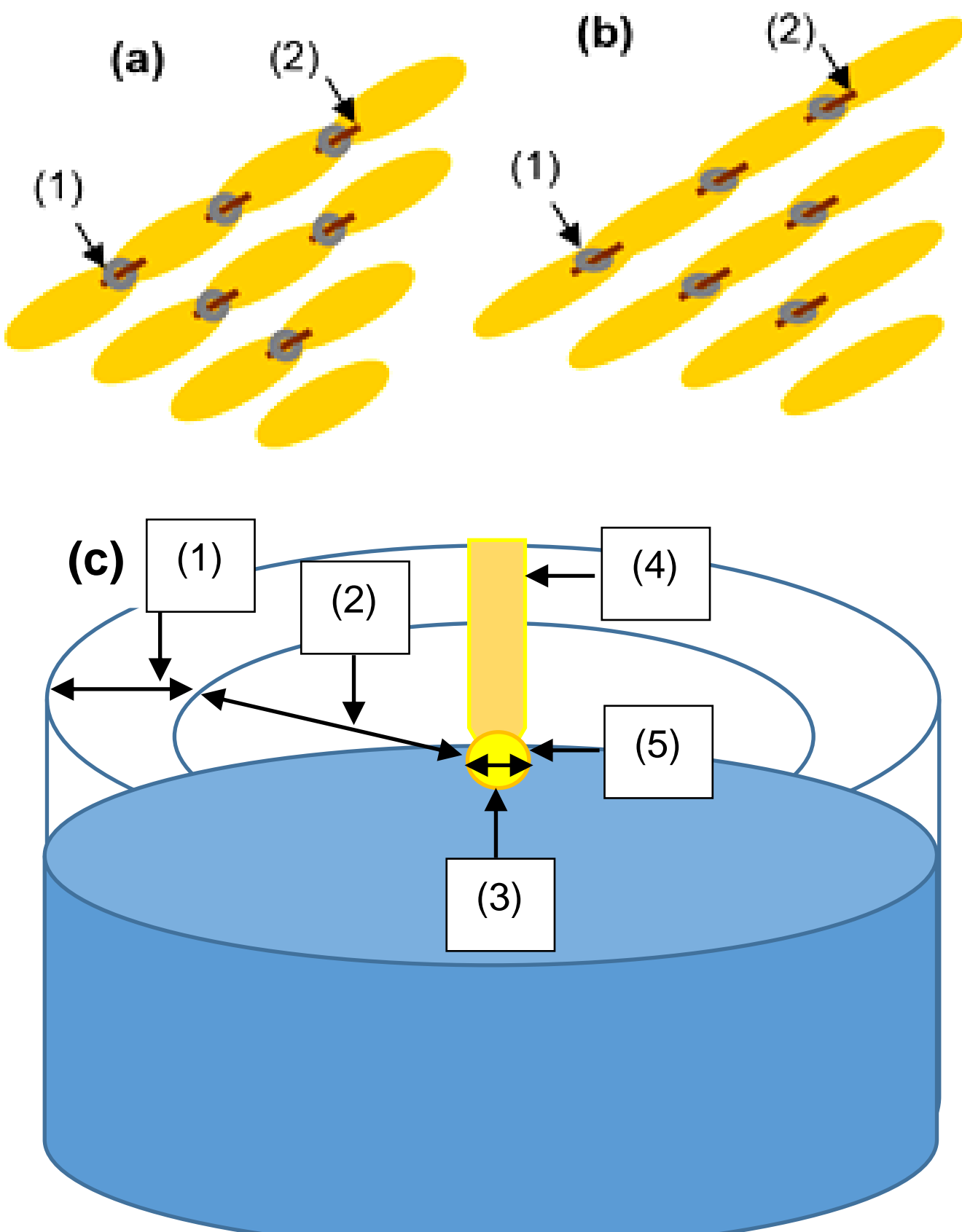

**Figure 4:** A triangular-shaped tiny particle showing (a) elongation of the array atoms at a lower rate and (b) elongation of the array atoms at a higher rate; (1) an energy knot in the lower-positioned atom and (2) an electron in the upper-positioned atom. (c) Solution beaker showing (1) electronically flat solution surface, (2) electronically decreasing solution surface, (3) center of concave meniscus, (4) tube carrying the glow sources, and (5) glow spot (Sketches drawn in estimation)

In the liquid state, atoms deal with less elongation. In the re-crystallization state, atoms have less elongation. The elongation rate of the atoms depends on the degree of hotness of the solution. The arrays of atoms convert into structures of smooth elements. Atoms of arrays are more like lines in shapes. It is called a structure of the smooth element. (The word 'element' does not relate to the periodic table element.) In the atomic elongation, a tiny particle keeps its triangular shape. In the development of smooth element structures, a triangular-shaped tiny particle is one unit.

### *3.4. Photon-matter interactions at the air-solution or air-matter interface*

When photons, having characteristics of the current, leave the splitting argon atoms in the pulse-based process, as discussed elsewhere [10], their 'force energy' decreased in the wavelength of the visible range. When those photons travel along the air-solution interface, their 'force energy' influences the matter at the solution surface. The matter, in the size of a tiny particle, keeps the elongated atoms. Those elongated atoms keep perturbed state electrons. Photons travel along the interface. They influence those perturbed state electrons.

The elongated atoms, having the disoriented electrons, belong to the arrays. The arrays belong to a triangular-shaped tiny particle. Through the field of force and the heat of energy, sections of the crest and trough of the traveling photons positioned the disoriented electrons of the atoms. In a triangular-shaped tiny particle, the structures of the smooth elements further flatten. It is under field permeation and heat dissipation. In each array, the disoriented electrons of the elongated atoms align. They align from both sides of the array. Arrays of elongated atoms form the structures of smooth elements in a triangular-shaped tiny particle. Figure 5 (a) shows this.

Photons traveling along the air-solution interface flatten the structures of smooth elements from both sides. Two structures of smooth elements maintain the constant gap. It is the inter-spacing distance. Figure 5 (a) also shows this. In another model, two atoms adhered in a slightly tilted position relative to each other. Figure 5 (b) shows this. However, traveling photons with suitable force-energy align the tilted atoms. The featured photons travel parallel to the air-solution interface. They align the positions of those atoms by tuning them. A smooth element structure has a line shape. There is an alignment of the perturbed state electrons of the elongated atoms.

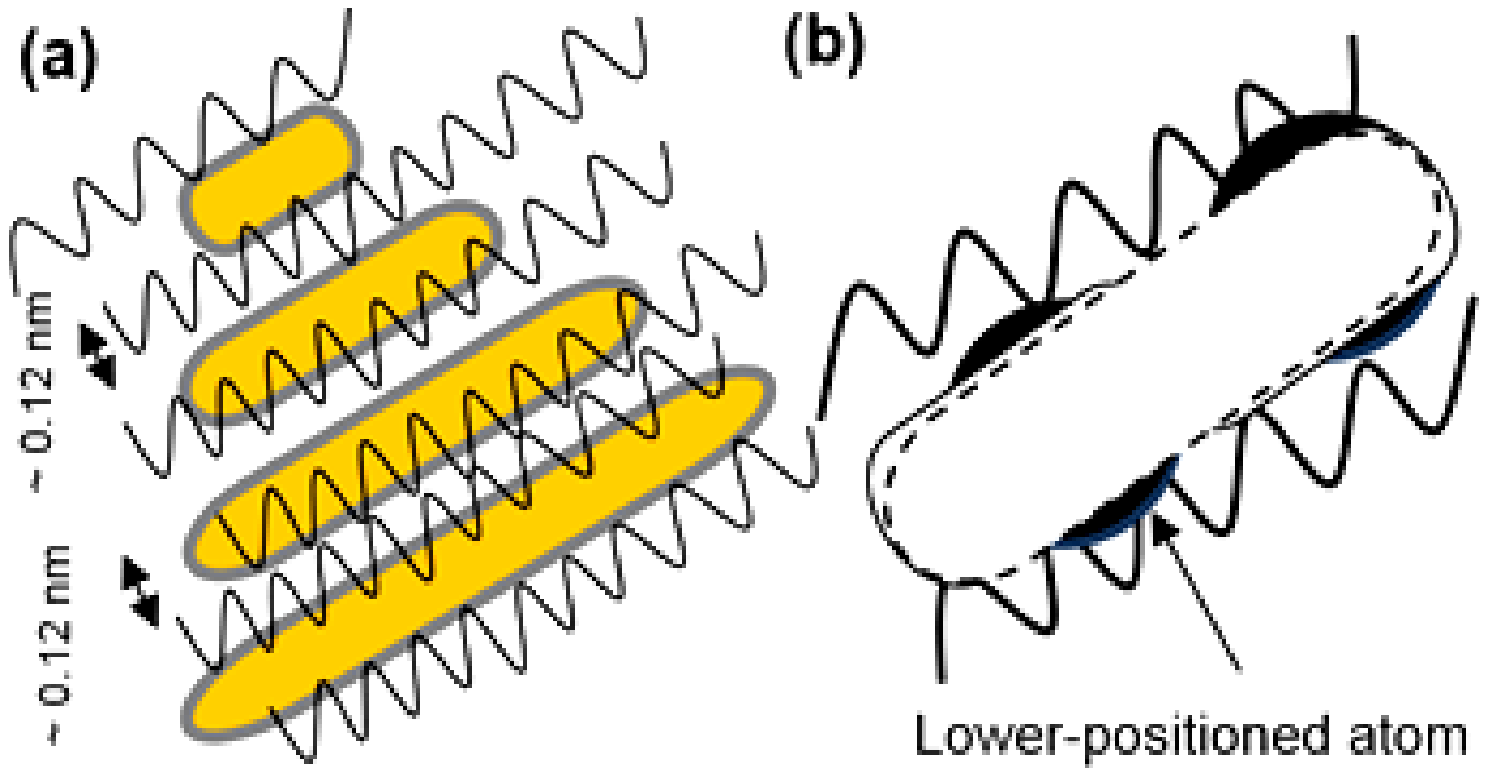

**Figure 5:** (a) A triangular-shaped tiny particle further sharpens the structures of smooth elements upon experiencing traveling photons, and (b) traveling photons along the air-solution interface, aligning the perturbed state electrons of slightly tilted atoms to convert them into the perfect structure of the smooth element

The triangular-shaped tiny particles enable the development of highly geometric particles. The studies elsewhere [2-5, 9, 32] discuss the highly geometric particles. The width of the smooth element structure in those different-shaped particles is approximately 0.12 nm. A study elsewhere [32] shows the actual width, which is also around 0.12 nm. In Figure 5 (a), the approximate width of the smooth element structure is also 0.12 nm. Each smooth structure of the atomic array in a triangular-shaped tiny particle maintains a constant width. The distance between the smooth element structures relates to the inter-spacing distance.

The widths of the smooth element structure and the inter-spacing distance in Figure 5 are the same. Depending on the application, the width of the smooth element structure can vary. The width of the structure of a smooth element can vary with the pulse rate, as discussed elsewhere [4]. The width of the structure of a smooth element can also vary under the application of force exerted in space and grounded formats [5].

Photons travel along the air-solution (or air-matter) interface. Photons experience different interactions with matter. They do not trap while traveling along the interface. The trapping of photons by the atomic lattices of a tiny particle is not the case. In the presence of surface ligands or capping agents, the travelling photons have a lesser impact on the atoms of a tiny particle. In the presence of surface ligands and capping agents, the photon influence on the atoms (of a tiny particle) needs further work. The atoms of the tiny particle flatten due to the force-energy of traveling photons. Those photons, which did not travel parallel to the interface, deteriorated the atomic lattices.

The atoms (of a tiny particle) do not oscillate collectively. The traveling photons change their characteristics and functions. The propagation of photons occurs through the aligned interstate electron gaps of the atoms of a tiny particle [10]. The studies elsewhere [5, 6] also did not validate the surface plasmon phenomenon. As discussed elsewhere [12], the surface plasmon phenomenon occurs in a single-layer tiny particle. In this case, a structure evolves. The photon influence converts atoms of each array into a line shape. The parallel-traveling photons can also influence the electronic orientation. However, more detail is needed to develop a better understanding.

### *3.5. Mono-layer shape dealing with gravity and levity at the concave meniscus*

Triangular-shaped tiny particles developed in the different regions of the electronically flat solution surface. Due to the force of immersion, they enter the electronically decreasing solution surface. Triangular-shaped tiny particles reach the region of the concave meniscus from different directions. Depending on their number and orientation, they shape a mono-layer in the region of the concave meniscus. The arrays of elongated atoms (or structures of smooth elements) adhere to each other. Their adherence occurs at the common center, which is also the center of the concave meniscus.

In the second stage, a mono-layer extends the shape. In this stage, smooth element structures belong to the later coalesced triangular-shaped tiny particles. Depending on the shape, the underlying science of the nucleating mono-layer also varies. To develop the mono-layer, several factors contribute. When the mono-layer has six sides, it develops a hexagonal shape. Figure 6 (a) shows this. Another mono-layer with a similar shape develops at the same point. An earlier-developed mono-layer automatically steps down. An earlier-developed mono-layer left the space above for a later-developed mono-layer. The force of gravity in a later-developed mono-layer is greater than the force of levity.

An earlier-developed mono-layer keeps a greater levitational force than the gravitational force. In the earlier-developed mono-layer, the gravitational force largely disappears. It is due to the upper mono-layer. An earlier-developed mono-layer is one step down from the solution surface. A 'one step down' is up to the lateral atomic width. The adherence of the mono-layers develops a nanoparticle or particle.

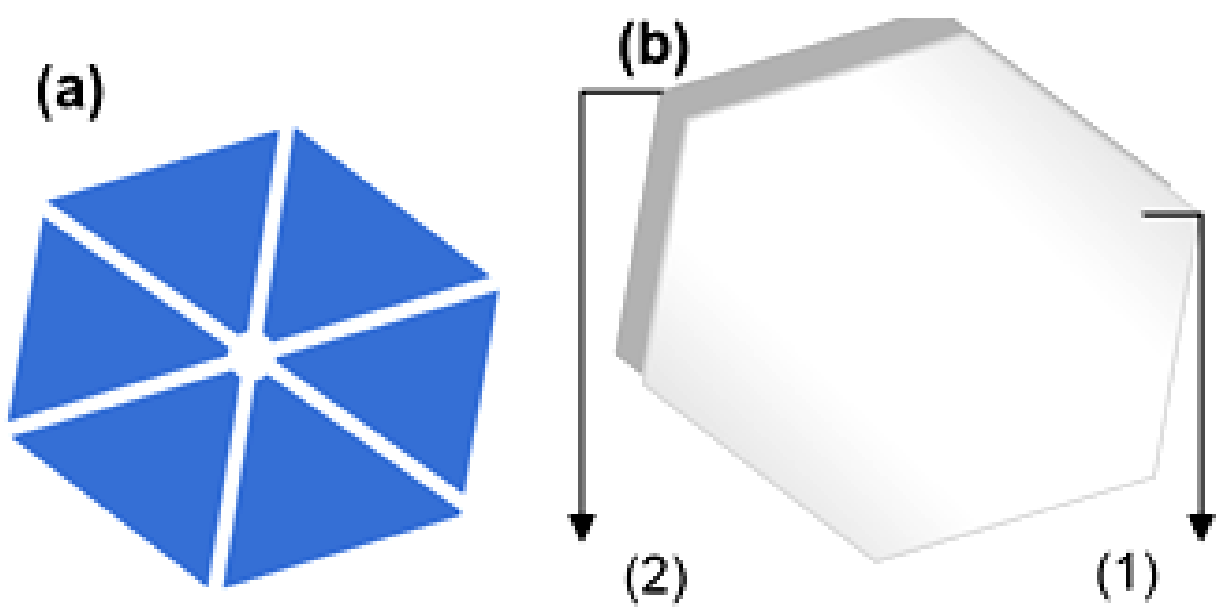

**Figure 6:** (a) A mono-layer shape similar to a hexagon. (b) Adherence of two mono-layers: (1) above-developed mono-layer and (2) below-developed mono-layer

Two mono-layer shapes adhere (or bind) laterally. The time to bind two mono-layer shapes is only a few microseconds. On exceeding the time to develop the third mono-layer, two laterally-adhered mono-layers will sink. A single mono-layer can stay at the solution surface. The gravity in that mono-layer will remain equal to the levity. Four forces remain in exertion during the format of an immersing force, as discussed elsewhere [32]. In the pulse-based process, a glow (containing photons and electron density) enters the solution. The center of the meniscus is also the center of the glow. A study elsewhere [32] also discussed this.

The earlier-developed mono-layer reaches one step down in developing another mono-layer. Thus, the earlier-developed mono-layer fills the space equal to the lateral atomic width. The later-developed mono-layer has a greater gravitational force than the levitational force. The below-positioned mono-layer considerably stops the levitational force ($F_L$) for the above-positioned mono-layer. In this way, the above-positioned mono-layer shape has a greater gravitational force ($F_G$) than its $F_L$. Equation (1) shows this.

**$F_G$ of above-positioned mono-layer > $F_L$ of above-positioned mono-layer … (1)**

The above-positioned mono-layer considerably stops the $F_G$ for the below-positioned mono-layer. In this way, the below-positioned mono-layer has more '$F_L$' than '$F_G$'. Equation (2) shows this.

**$F_L$ of below-positioned mono-layer > $F_G$ of below-positioned mono-layer … (2)**

Due to the greater $F_L$ in the earlier-developed mono-layer shape and the greater $F_G$ in the later-developed mono-layer shape, they adhere (or bind) laterally. In developing a hexagonal-shaped particle, the above mono-layer shape adheres to the below mono-layer shape. Figure 6 (b) shows this. The whole process of adhering to two mono-layers took only a few microseconds. Understanding the gravity versus levity (or the levity versus gravity) is an important factor. Do the properties of the tiny particles influence the mono-layer shape? It is also an important factor.

### *3.6. General discussion*

Controlling the shape and structure of nanoparticles and particles depends on the pulse energy, as discussed elsewhere [4]. Tuned pulses or other suitable means are sources of nanoenergy packets. Triangular-shaped tiny particles can develop under the supply of nanoenergy packets. The solid atoms keep the ground points below the ground surface [12]. It is not the case in developing triangular-shaped tiny particles. Different synthetic protocols selected in earlier studies developed a variety of nanoparticles and particles.

Suitable ligands and surfactants can also enable the development of triangular-shaped tiny particles. After dissociation from the precursor, and by decreasing the potential energy of the electrons, atoms undergo a re-crystallization state. Therefore, transitional atoms amalgamate at the solution surface. Atoms in suitable elements can reach the re-crystallization state at the solution surface. For this, they levitate. Many studies published in the literature explain the synthesis and application of particles. Those studies are discussed elsewhere [34-50]. However, the science behind the development of those particles needs further investigation.

This study is the beginning of colloidal science and physical chemistry. New research efforts should also consider the new atomic and electronic structures discussed elsewhere [8, 51]. When the atoms elongate suitably, their triangular-shaped tiny particles can develop mono-layer shapes, nanoparticles, or particles. It is feasible to size a nano-object or particle. Using different capping agents cannot protect the atoms from transitional behavior. Atoms of the suitable elements retain their original shape. It is the case in the structural evolution. A study elsewhere [12] discussed this.

A smooth element structure has the modified state atoms. The adherence of structures of smooth elements occurs in an adjacent manner. A mono-layer shape develops. From outside their zones, atoms in suitable elements do modify their structures. It is usually the case under overwhelming processing or synthesis conditions of their materials. A study elsewhere [52] discusses the structural modification of an atom.

How do photons smooth the structures of the elements of a mono-layer shape developing at the air-solution interface? It needs further investigation. Key factors in the adherence of two mono-layers within microseconds need to be studied further. In the adherence process, how do gravity and levity play a role at a more granular level? It also needs to investigate. Different microscopic and X-ray studies can further examine mono-layer shapes and particles.

In some industrial applications, such as catalysis, sensors, and coatings, the mono-layer shapes and particles can outperform traditional materials. Theoretical models or simulations can predict the behavior of triangular particles under different conditions. Computational models can explore the stability and dynamics of mono-layer adhesion more comprehensively. The mono-layers and their particles also show a great potential in sensitive fields. In the healthcare and advanced technology fields, they are also valuable.

There is a need to investigate the chemical composition, mechanical properties, and surface roughness by energy-dispersive X-ray spectroscopy and atomic force microscopy. To study photon alignment and the force field is also important. The force contribution to develop the mono-layer shape requires more work. The study of nanoenergy packets and photon-matter interactions also requires more work. The theoretical and mathematical formulations related to the conceptual description of interatomic binding will further broaden the field.

## 4.0. Conclusion

In processing different colloidal solutions, atoms dissociate from their precursor. It is the first step. Atoms levitate to form a monolayered assembly at the solution surface. It is the second step. Atoms attain the required transition state while levitating. The atoms of transitional behaviors stick to the solution surface. Atoms float in a new environment of the air-solution interface. A monolayered assembly develops at an electronically flat solution surface. Binding in the atoms is due to a pulse energy or another energy source.

A block of two tiny particles develops. It has the shape of a binding energy or nanoenergy packet. Each tiny particle keeps a triangular shape. Connected triangular-shaped tiny particles separate due to a difference in force at the electron level. The force difference exists between the electronically flat solution and the electronically decreasing solution surfaces. After separation, tiny particles address individual dynamics.

Arrays of triangular-shaped tiny particles enter the electronically decreasing solution surface. The tilting of the electrons in their atoms is adjacent-wise. Electrons of elongating atoms maintain their energy and force owing to the established potential and orientation. The side-to-side positioned atoms in each array stretch their lattices unidirectionally. By keeping the position, the atoms of each array elongate at a more or less constant rate. The elongation rate of an atom depends on the conditions of the solution processing.

From both sides of the array, the parallel traveling photons align the modified atoms by tilting their electrons. Permeation of the force and dissipation of the heat of the parallel-travelling photons align the misaligned electrons. The widths of a smooth element structure and the inter-spacing distance are the same. If there is any discrepancy in the position of the modified atoms, photons travel parallel to the interface position and align them. The plasmonic phenomenon does not reveal the collective oscillation of the atoms of a tiny particle. In this study, it is a field-permeating heat-dissipating phenomenon rather than a surface plasmon phenomenon.

In a solution-based process, triangular-shaped tiny particles arrive at the region of the concave meniscus. Arrays of atoms convert into the structures of smooth elements. In the meniscus region, a mono-layer shape develops. The development is from the adherence of the smooth element structures adjacent-wise. The structures of smooth elements in the following tiny particles further extend a mono-layer shape. There is a development of another (above) mono-layer shape in the meantime.

The gravitational force of the upper mono-layer shape is greater than the levitational force. For a lower mono-layer shape, the force at the electron level operates in opposite functions in nature. In the lateral-wise binding of mono-layers, the force at the electron level acts locally. In a few microseconds, two mono-layer shapes adhere laterally.

The model study discussed here shows that it is feasible to develop devices in different size ranges. This work can also open new theoretical, simulation, computation, and design fields. The roles of gravity and levity in the upper and lower mono-layers demand more detail. How do these forces interact with the surface tension and other interfacial forces during mono-layer formation? The mathematical models or simulations to quantify these forces would add depth. There is also a need to assess the long-term stability of the mono-layers and their particles. The conditions, such as temperature, humidity, or exposure to different chemical agents, also need to be investigated.

The author would further recommend discussions on reduction rates and capping agents. How does the rate of precursor reduction impact the formation of tiny-shaped particles in the proposed model? Could surface ligands or capping molecules modify the photon-matter interactions discussed in this study? There is an acute need to address these questions. It is also necessary to establish the mathematical formulations. They will describe the relationship between the forces and the atomic assembly process.

**Funding Declaration**:

There is no funding.

**Conflicts of interest:**

The author declares no conflicts of interest.

**Data availability statement:**

The existing study does not contain any data.

## Bibliographic detail:

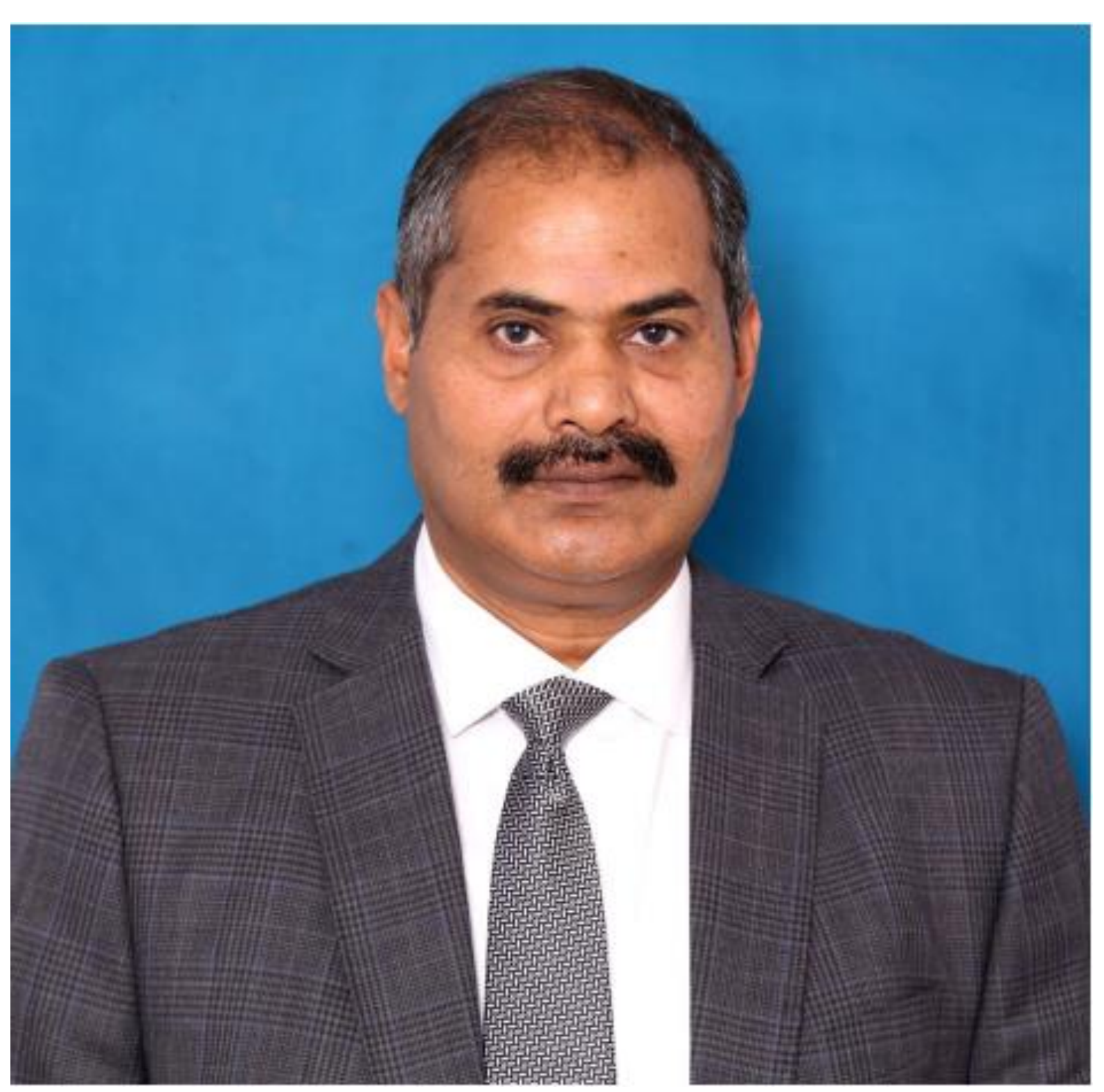

In the year 1996, **Mubarak Ali** obtained a B.Sc. degree in Physics and Mathematics. It was from the University of the Punjab. Mubarak Ali earned the M.Sc. degree in Materials Science in 1998. Mubarak Ali completed the MSc degree with distinction from the Bahauddin Zakariya University, Multan. He completed the MSc thesis from the Quaid-i-Azam University, Islamabad. From the Universiti Teknologi Malaysia, Mubarak Ali earned a PhD degree in Mechanical

Engineering (Major in Materials Engineering). The Malaysian government awarded a PhD scholarship under the Malaysian Technical Cooperation Programme, MTCP, in 2004-07. In 2010, the Scientific and Technological Research Council of Turkey (TÜBİTAK) funded the postdoctoral research under the foreign fellowship programme. Dr Mubarak earned the first postdoctorate in advanced surface technologies from the Department of Metallurgical and Materials Engineering at Istanbul Technical University. Dr Mubarak completed another postdoc in nanoscience and nanotechnology from Tamkang University (Physics Department), Taipei, in 2013-2014. The National Science Council, now the Ministry of Science and Technology, Taiwan, sponsored his second postdoctoral studies. He worked as an Assistant Professor on the tenure track at COMSATS University Islamabad from May 2008 to June 2018 (previously known as the COMSATS Institute of Information Technology). His new position is in process. He also worked as an assistant director and deputy director in the Pakistan Council of Renewable Energy Technologies, Ministry of Science and Technology, Islamabad, from January 2000 to May 2008. The Institute for Materials Research at Tohoku University, Japan, invited Dr Mubarak to deliver a scientific talk. He presented scientific research at many conferences and in many countries. The main research areas include basic science, physics, materials science, surface science, surface engineering, materials chemistry, physical chemistry, energy science, nanoscience, and sustainability. Dr Mubarak also won a merit scholarship for PhD studies from the Higher Education Commission, Pakistan. Due to another position, he did not obtain this opportunity. In the year 2000, Mubarak earned a diploma in English language from the National University of Modern Languages, Islamabad. From the same institution, Dr Mubarak Ali completed the Japanese language certificate in 2001. Dr Mubarak Ali is the author of several articles given at the links https://www.researchgate.net/profile/Mubarak_Ali5 & https://scholar.google.com.pk/citations?hl=en&user=UYjvhDwAAAAJ